\documentclass[english,showpacs,hyperref,superscriptaddress,twocolumn]{revtex4-1}
\usepackage[T1]{fontenc}
\usepackage[latin9]{inputenc}
\usepackage{babel}
\usepackage{textcomp}
\usepackage{mathrsfs}
\usepackage{amsmath}
\usepackage{amssymb}
\usepackage{graphicx}
\usepackage[unicode=true,pdfusetitle,
 bookmarks=true,bookmarksnumbered=false,bookmarksopen=false,
 breaklinks=false,pdfborder={0 0 1},backref=false,colorlinks=false]
 {hyperref}
\usepackage{breakurl}

\makeatletter
\@ifundefined{textcolor}{}
{
 \definecolor{BLACK}{gray}{0}
 \definecolor{WHITE}{gray}{1}
 \definecolor{RED}{rgb}{1,0,0}
 \definecolor{GREEN}{rgb}{0,1,0}
 \definecolor{BLUE}{rgb}{0,0,1}
 \definecolor{CYAN}{cmyk}{1,0,0,0}
 \definecolor{MAGENTA}{cmyk}{0,1,0,0}
 \definecolor{YELLOW}{cmyk}{0,0,1,0}
 }

\usepackage{amsfonts}\usepackage{subfigure}

\providecommand{\U}[1]{\protect\rule{.1in}{.1in}}

\makeatother

\begin{document}

\title{Theoretical and experimental aspects of quantum discord and related measures}

\author{Lucas C. C\'{e}leri}

\email{lucas.celeri@ufabc.edu.br}

\selectlanguage{english}

\affiliation{Centro de Ci\^{e}ncias Naturais e Humanas, Universidade Federal do ABC, R. Santa Ad\'{e}lia 166, 09210-170 Santo Andr\'{e}, S\~{a}o Paulo, Brazil}

\author{Jonas Maziero}

\email{jonas.maziero@ufabc.edu.br}

\selectlanguage{english}

\affiliation{Centro de Ci\^{e}ncias Naturais e Humanas, Universidade Federal do ABC, R. Santa Ad\'{e}lia 166, 09210-170 Santo Andr\'{e}, S\~{a}o Paulo, Brazil}

\author{Roberto M. Serra}

\email{serra@ufabc.edu.br}

\selectlanguage{english}

\affiliation{Centro de Ci\^{e}ncias Naturais e Humanas, Universidade Federal do ABC, R. Santa Ad\'{e}lia 166, 09210-170 Santo Andr\'{e}, S\~{a}o Paulo, Brazil}

\begin{abstract}
Correlations are a very important tool in the study of multipartite systems, both for classical and quantum ones. The discussion about the quantum nature of correlations permeates Physics since Einstein, Podolski and Rosen published their famous article criticizing quantum mechanics. Here we provide a short review about the quantum nature of correlations, discussing both its theoretical and experimental aspects. We focus on quantum discord and related measures. After discussing their fundamental aspects (theoretically and experimentally), we proceed by analysing the dynamical behaviour of correlations under decoherence as well as some applications in different scenarios, such as quantum computation and relativity, passing through critical and biological systems.
\end{abstract}

\maketitle

\section{Introduction}

Correlations are ubiquitous in nature and have been playing an important role in human life for a long time. For instance, in economy correlations between price and demand are extremely important for a businessman (or even for a government) to decide his investment policy. In the field of biology, the genetic correlations are fundamental to follow individual traits. The relationship between income distribution and crime rate is just one example coming from the Social Sciences. Broadly speaking, correlation is a quantity that describes the degree of relationship between two variables. In the classical domain, such a quantity can be measured within the framework of information theory, developed by Shannon in 1948 \cite{Shannon}.

Let us consider first the classical scenario with two distinct random variables, $A$ and $B$, with well defined probability distributions $\left\{p_{A}\right\} $ and $\left\{p_{B}\right\} $, respectively. We can think of $A$ and $B$ in terms of probability sets for the outcome of measurements performed on the joint system or on the two different subsystems, usually named \textit{Alice} and \textit{Bob}. To say that $A$ and $B$ are correlated in some form means that the joint probability distribution $\left\{p_{AB}\right\} $ cannot be written in the product form $\left\{p_{A}\times p_{B}\right\} $. To quantify the amount of correlations shared by the two classical random variables, Shannon introduced the mutual information \cite{Shannon}:
\begin{equation}
I\left(A:B\right)\equiv H\left(A\right)+H\left(B\right)-H\left(A,B\right),
\label{MInformation}
\end{equation}
with $H\left(X\right)\equiv-\sum_{x\in\mathcal{X}}p_{x}\log p_{x}$ being the well-known Shannon entropy. Here and throughout this article we use the short-hand notation $p_{x}\equiv p_{X=x}$, in which the random variable $X$ assumes the value $x$ from the set $\mathcal{X}$, and we use $\log$ to represent the binary logarithm. So, the correlations are quantified in bits.

In this article we are interested in a very special kind of correlation, the ones arising from the superposition principle underlying quantum mechanics, known as \emph{non-classical} (or quantum) correlations. As we shall see, this is a very intricate issue for which we do not have a closed theory yet. First we observe that the simplicity and generality of Eq. (\ref{MInformation}) allows its direct extension to the quantum domain as a measure of the total correlations in a bipartite system, 
\begin{equation}
\mathcal{I}\left(A:B\right)\equiv S\left(A\right)+S\left(B\right)-S\left(A,B\right),
\label{QMI}
\end{equation}
where $S\left(X\right)\equiv-\mbox{Tr}\rho_{X}\mbox{log}\rho_{X}$ is the von Neumann entropy \cite{Vedral1}. The reduced density matrix of partition $X$, $\rho_{X}$, is obtained from the entire density operator by means of the partial trace operation $\rho_{X}\equiv\mbox{Tr}_{Y}(\rho_{XY})$. However, differently from the classical scenario, here we face the problem of distinguishing between the various types of correlations. As we shall see below, such problem does not have a clear-cut solution hitherto.

Since the early years of quantum mechanics, \emph{non-local} correlations have been at the core of a long standing debate about the foundations of the theory. We may say that the discussion started in 1935, when Einstein, Podolsky and Rosen (EPR) published their famous article entitled \emph{{}``Can quantum-mechanical description of physical reality be considered complete?\textquotedblright{}} \cite{EPR}. They considered two spatially separated particles, $A$ and $B$, having both perfectly correlated positions and momenta. By defining a complete theory as the one for which there is one element of physical reality corresponding to each element of the theory and assuming \emph{local realism}, i.e., shortly
\begin{itemize}
\item \emph{locality} - There is no action at a distance;
\item \emph{realism} - {}``If, without disturbing a system, we can predict with certainty the value of a physical quantity, then there exists an element of physical reality corresponding to this physical quantity.\textquotedblright{}; 
\end{itemize}
they argued that quantum mechanics would not be complete (see \cite{Leuchs} for a recent review about the EPR paradox). In a sense, EPR implied that quantum mechanics should be extended, possibly through the introduction of hidden variables, making the new theory compatible with their premise of local realism stated above. In other words, EPR demonstrated an inconsistency between local realism and the completeness of quantum mechanics.

Schr\"{o}dinger noted that at the core of the paradox lies the very structure of the Hilbert space \cite{Sch}. He pointed out that the state vector of the entire system (including both particles $A$ and $B$) is \emph{entangled}. This means that we cannot ascribe an individual state vector for each particle separately, only the entire system can have such a mathematical representation. This fact leads to the definition of entanglement. A pure state is separable if and only if it can be written in the product form: 
\begin{equation}
\vert\psi_{AB}\rangle=\vert\psi_{A}\rangle\otimes\vert\psi_{B}\rangle,
\label{ent}
\end{equation}
where $\vert\psi_{A}\rangle$ ($\vert\psi_{B}\rangle$) is the state vector describing the system $A$ ($B$) alone. Otherwise it is entangled. The argument of Schr\"{o}dinger against EPR was that this entanglement is degraded during the process by which $A$ and $B$ become spatially separated \cite{Sch}, leading to the impossibility of physically realizing the \emph{Gedankenexperiment} proposed by EPR. Although interesting, this argument does not settle the paradox.

Since then, entanglement has been at the core of the discussion about the non-local aspects of quantum theory, both in respect to its fundamental aspects as well as for (more recently) technological applications. The efforts that lead us to the theory of entanglement have been mainly focused on two aspects of the problem: Given a certain state, is it entangled? If yes, how much entanglement is there? Besides many advances (both in the theoretical and in the experimental fields) have been achieved, a complete theory describing entanglement is still lacking. We refer the reader to the reviews in Ref. \cite{Entanglement} for a more complete discussion about entanglement. Before we proceed, let us define the entanglement for more general states, the mixed states. For systems described by a density operator $\rho_{AB}$, the separability condition given in Eq. (\ref{ent}) becomes 
\begin{equation}
\rho_{AB}=\sum_{i}p_{i}\rho_{A}^{i}\otimes\rho_{B}^{i},
\label{entmix}
\end{equation}
with $\{p_{i}\}$ being a probability distribution and $\rho_{X}^{i}$ a density operator for the subsystem $X$. In the case of continuous variables, we must replace the summation by an integral over the entire probability space. The state (\ref{entmix}) is the most general bipartite state that Alice and Bob can create using local quantum operations and classical communication (LOCC). Therefore, an entangled state is that one that cannot be prepared by LOCC.

Let us now go back to EPR. Despite all the progress achieved in the path to understanding the EPR paradox, it was only in 1964 that it was put on a mathematical ground, making possible its experimental investigation. Considering a model described by local hidden variables (LHV), Bell proved a theorem, today consisting in the so-called Bell inequalities, confronting the predictions of quantum mechanics with those coming from the local realistic assumption \cite{Bell}. If a given state violates a Bell inequality then, it is not possible to describe such a state by means of an LHV model, implying that local realism must be abandoned. A recent review about the theoretical and the experimental aspects of this theorem can be found in Ref. \cite{Leuchs}. We observe here that the extension of the quantum theory proposed by Bell is classical, in the sense that the hidden variables are local in nature. However, more recently, Colbeck and Renner have shown, under the assumption that the measurement setting can be freely chosen by the observer, that the quantum theory cannot be extended at all \cite{Colbeck}.

With concern to correlations, the scenario regarding what was discussed until here is the following: The non-classical nature of correlations was attributed to states that violate some Bell inequality. This kind of correlation, also named non-local, cannot be explained by an LHV model. However, this situation changed in 1989, when Werner published an article showing that there are mixed entangled states that do not violate any Bell inequality and therefore could be described by means of an LHV model \cite{Werner}. This was a surprising fact because it indicates that there are non-separable (entangled) mixed states that could be described in {}``classical terms\textquotedblright{} by an LHV model. So, another kind of correlation arises, the one contained in an entangled state, but that do not violate any Bell inequality. This kind of correlation is indeed non-classical since we disregard LHV models. This picture remained unchanged until 2001, when it was identified a new kind of quantum correlation, which may be present even in separable (non-entangled) states.

\section{Quantum Discord}

This new scenario emerged from an nonequivalence between the classical and the quantum versions of information theory. We may explore this nonequivalence in several ways in order to reveal some non-classicality. In what follows we will present a way to do it that we consider very illustrative. From Bayes' rule we can write the classical conditional probability for obtaining the value $a$ for the random variable $A$ when the value of the random variable $B$ is known to be $b$, $p_{a|b}$, in the form $p_{a|b}\equiv p_{a,b}/p_{b},$ which leads to an equivalent expression for the classical mutual information (\ref{MInformation})
\begin{equation}
J(A:B)\equiv H(A)-H(A|B),
\label{CMI2}
\end{equation}
where the classical conditional entropy reads $H(A|B)\equiv-\sum_{a,b}p_{a,b}\log p_{a|b}=H(A,B)-H(B)$.

While Eq. (\ref{MInformation}) has a direct extension to the quantum territory, as shown in Eq. (\ref{QMI}), here we are faced with a fundamental fact: A measurement generally disturbs a quantum system. The problem arises when we try to extend the classical concept of the conditional entropy $H(A\vert B)$ to the quantum domain. This quantity measures the uncertainty about the random variable $A$ after we \emph{measure} $B$. Therefore, its extension to quantum mechanics presents some ambiguity due to the fact that, depending on the observable we choose to measure $B$, the value of the conditional entropy would be different and may even be negative.

In order to obtain a quantum version of Eq. (\ref{CMI2}), let us consider a set of measurements $\Pi_{j}^{(B)}$ on subsystem $B$ of the composite state $\rho_{AB}$. The reduced state of subsystem $A$, after the measurement, is given by 
\begin{equation}
\rho_{A}^{j}=\frac{1}{q_{j}}\mbox{Tr}_{B}\left\{ \left(\mathbf{1}_{A}\otimes\Pi_{j}^{(B)}\right)\rho_{AB}\left(\mathbf{1}_{A}\otimes\Pi_{j}^{(B)}\right)\right\} ,
\label{CondState}
\end{equation}
where $q_{j}=\mbox{Tr}_{AB}\left\{\left(\mathbf{1}_{A}\otimes\Pi_{j}^{(B)}\right)\rho_{AB}\right\} $ is the probability for the measurement of the $j$-th state in subsystem $B$ and $\mathbf{1}_{A}$ is the identity operator for subsystem $A$. For a complete set of measurements $\left\{\Pi_{j}^{(B)}\right\} $, we can define the conditional entropy of subsystem $A$, for a known subsystem $B$, as 
\begin{equation}
S\left(A|B\right)\equiv\sum_{j}q_{j}S\left(\rho_{A}^{j}\right).
\label{AverCondEnt}
\end{equation}
Thus, we can also define the following quantum extension for Eq. (\ref{CMI2}) 
\begin{equation}
\mathcal{J}(A:B)=S(A)-S\left(A|B\right).
\label{QMI2}
\end{equation}

While in the classical case we have the equivalence between both definitions of the mutual information in Eqs. (\ref{MInformation}) and (\ref{CMI2}), i.e., $I-J=0$, for a quantum state, Eqs. (\ref{QMI}) and (\ref{QMI2}) are not equivalent in general any more. The difference, 
\begin{equation}
\mathcal{D}(A:B)\equiv\mathcal{I}(A:B)-\max_{\left\{ \Pi_{j}^{(B)}\right\} }\mathcal{J}(A:B),
\label{Dis}
\end{equation}
was called \emph{quantum discord} (QD) by Ollivier and Zurek \cite{OllZur}. One can say that Eq. (\ref{Dis}) reveals the quantumness of the correlations between partitions $A$ and $B$, since it shows the departure between the quantum and the classical versions of information theory. An important observation is that QD captures the non-classical aspects of correlations contained in certain states, which includes entanglement. However, while an entanglement measure \cite{Entanglement} vanishes for a separable state, QD may be non-zero for states of the form (\ref{entmix}). For the simplest case of bipartite pure states, QD is equal to entanglement.

The information theoretic approach to study correlations also leaded us to a new definition of classical correlations. A quantum composite state may have the support of classical correlations, $\mathcal{C}(A:B)$, which, for bipartite quantum states, can be quantified via the measure proposed by Henderson and Vedral \cite{HenVed}: 
\begin{equation}
\mathcal{C}(A:B)\equiv\underset{\left\{ \Pi_{j}^{(B)}\right\} }{\max}\left[S(A)-S(A|B)\right],
\label{HV}
\end{equation}
where the maximum is taken over the complete set of positive operator valued measurements (POVM) $\left\{ \Pi_{j}^{(B)}\right\}$ on subsystem $B$. For two-qubit systems, since the conditional entropy is concave over the set of POVMs (which is convex), the optimization is performed on the extreme points of the set of POVMs, which are rank 1 \cite{Devetak}. For more general systems, this optimization can be performed over the complete POVM set, which is a very difficult problem, even numerically.

From the above definitions, it follows immediately that $\mathcal{D}(A:B)+\mathcal{C}(A:B)=\mathcal{I}(A:B)$. For pure states, we have a special situation where the quantum discord is equal to the entropy of entanglement and also equal to the Henderson-Vedral classical correlation. In other words, $\mathcal{D}(A:B)=\mathcal{C}(A:B)=\left.\mathcal{I}(A:B)\right/2$ \cite{HenVed,Groisman}. In this case, the total amount of non-classical correlation is captured by an entanglement measure. On the other hand, for mixed states, the entanglement is only a part of the non-classical aspects of correlations \cite{OllZur,White,Caves}. We may visualize the different {}``flavors\textquotedblright{} of non-classical correlations considering the Bell-diagonal class of states, as depicted in Fig. 1.

\begin{center}
\begin{figure}[th]
\centering{}\includegraphics[bb=0bp 0bp 7.5273000000000003in 4.7297000000000002in,width=5.2647in,height=5.0574in,keepaspectratio]{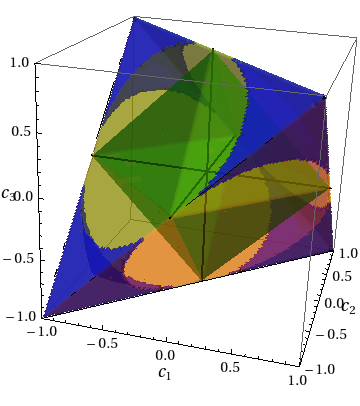}
\caption{Geometric representation of the different types of correlations in parameter space of Bell-diagonal states $\rho_{bd}=\left(\mathbf{1}_{AB}+\sum_{i=1}^{3}c_{i}\sigma_{i}^{A}\otimes\sigma_{i}^{B}\right)/4$, where $\left\{ \sigma_{i}^{A}\right\} _{i=1}^{3}$ are the usual Pauli matrices. Valid states are in the (yellow) tetrahedron. Separable states are those in the (green) octahedron, while classical states comprise the (black) axis. States that violates the CHSH inequality are found in the blue regions \cite{Spengler}.}
\end{figure}
\par\end{center}

The discovery that bipartite mixed separable (non-entangled) states can have non-classical correlations \cite{OllZur,Nonlocal} has opened a new perspective in the study and comprehension of quantum aspects of its nature. A recent result, that almost all quantum states have a non vanishing quantum discord \cite{Ferraro}, shows up the relevance of studying these correlations. Such quantum correlations, that may be present in separable states, are conjectured to play a role in the quantum advantage of tasks in mixed state quantum information science \cite{White,Caves,Adesso}.

There are some important issues about quantification of non-classical correlations beyond entanglement. One of the most relevant to be noted at this point is the fact that almost all non-classicality measures, as QD, are based on extremization procedures that constitute a difficult problem, even numerically. Actually, analytical solutions for the QD were obtained only recently for a certain class of highly symmetrical states \cite{Lou,Sarandy,Maziero} (see also \cite{Ali,Lu,Adesso,ChenX}). Hence, an alternative, operational (without any extremization procedure) quantifier is rather desirable.

Our aim in this article is to discuss some theoretical and experimental aspects of non-classical correlations, and its classical counterpart. The remaining of the article is organized as follows. In Sec. \ref{PI} we discuss some physical interpretations of the quantum discord (the most popular measure of quantum correlations). As we shall see, the quantification of quantum correlations is still an open problem. Sec. \ref{MC} is devoted to present other proposed measures for such correlations, the relationship between them, their main features, and some analytical results. Besides the quantification problem, there is another important issue, the one regarding the experimental detection of such correlations and Sec. \ref{WC} goes in this direction, by addressing the problem of witnessing non-classical correlations. The action of the decoherence on these correlations is discussed in detail in Sec. \ref{DC}, showing very interesting dynamical aspects. In Sec. \ref{AC} we review the discussion about the role played by quantum correlations in several scenarios, including quantum information processing and critical systems, passing through biology and relativistic effects on these correlations. Beyond theoretical developments, the experimental investigation of these correlations and of its dynamics is discussed in Sec. \ref{EC}. A brief summary and our final discussions are presented in Sec. \ref{CR}.

\section{Physical interpretations of quantum correlations}
\label{PI}

Quantum Information Science (QIS) introduced a new way to think about quantum mechanics (see \cite{ITF} for a more profound discussion in this direction). As mentioned in the last section, this new perspective motivates the introduction of a measure to quantify quantum correlations, the quantum discord. QD has been largely studied in a great variety of situations and a physical interpretation of such a quantity is quite desirable for the very foundations of the quantum theory. Until now, we have two proposed interpretations for QD, one in thermodynamics \cite{Zurek} and the other \cite{Cavalcanti,Datta} from the information theoretic perspective through the state merging protocol.

\subsection{Thermodynamic interpretation of quantum discord}
\label{TID}

In an attempt to understand the limitations of thermodynamics, Maxwell introduced a character (today named Maxwell's demon) that permeates physics since then. Furthermore, Maxwell's demon plays a role also in information theory. In fact, thermodynamics and information theory present several links. Developments in this direction have being achieved since the Brillouin treatment of the informational entropy (due to Shannon) and the thermodynamic entropy (due to Boltzmann) on the same footing \cite{Brillouin}. An important relation between these two distinct sciences was explored by Landauer to finally {}``exorcise\textquotedblright{} Maxwell's demon through his information erasure theorem \cite{Landauer}. For a recent review about the physics of Maxwell's demons from the perspective of information theory see Ref. \cite{VlatkoR}.

Maxwell's demon is usually modelled as a classical information processing device, whose role is to implement a suitable action based on previous information acquired by means of a classical measurement. The final goal of the demon is to extract work from the environment. Zurek then asked if a quantum version of the demon would be more efficient in accomplishing this task than the classical one. This quantum demon was defined as a being that is able to measure non-local states and implement quantum conditional operations. Zurek verified that not only the quantum demon really could extract more work than the classical one, but also that the difference in the work extraction is given by the QD \cite{Zurek}.

The machine considered by Zurek consists of a system $\mathcal{S}$ correlated with a measurement apparatus $\mathcal{M}$, and an environment $\mathcal{E}$ at temperature $T$. The demon reads the outcome of $\mathcal{M}$ and then uses the acquired information to extract work from $\mathcal{E}$ by letting $\mathcal{S}$ to expand throughout the available phase (or Hilbert) space. The difference in the work extraction by the classical and the quantum demons is given by \cite{Zurek} 
\begin{equation}
\Delta W=k_{B}T\mathcal{D}(\rho_{\mathcal{SM}}).
\end{equation}
In this equation, the quantum discord is given by 
\begin{equation}
\mathcal{D}(\rho_{\mathcal{SM}})=\min_{\{\vert M_{k}\rangle\}}\left[S(\tilde{\rho}_{\mathcal{M}})+S(\tilde{\rho}_{\mathcal{S}|\mathcal{M}})\right]-S(\rho_{\mathcal{SM}})
\end{equation}
where $\{\vert M_{k}\rangle\}$ is the measurement basis in the apparatus state space and $k_{B}$ is the Boltzmann constant. The post-measurement entropies of the apparatus and system are given, respectively, by $S(\tilde{\rho}_{\mathcal{M}})=H(\{p_{k}\})$ and $S(\tilde{\rho}_{\mathcal{S}|\mathcal{M}})$, with states and (conditional) entropies defined analogously to Eq. (\ref{CondState}) and (\ref{AverCondEnt}). This result tells us that the efficiency of the demon is determined by the information about $\mathcal{S}$ that is accessible to the demon \cite{Zurek} and that it can acquire more information by means of a quantum measurement when compared with a classical one. This is a direct relation between a thermodynamic quantity (work) and an information theoretic one (quantum discord). In other words, the quantum discord is a measure of the advantage obtained by means of the quantum dynamics. In \cite{Brodutch} it was shown that two different versions of quantum discord determine the difference between the efficiencies of a Szilard\textquoteright{}s engine under different sets of restrictions. It is important to observe here that despite the quantum nature of the demon, his memory is still classical.

A closely related discussion was presented in Ref. \cite{Vlatko1}, where the authors considered the Landauer's erasure principle, which states that in order to erase the information stored in a given system, we must perform an amount of work that is proportional to the entropy of that system, in the presence of a quantum memory. Then, they showed that the difference between the work cost of the erasure procedure using a quantum and a classical memory is given by $k_{B}T\mbox{ln}2\mathcal{D}(\mathcal{S}:\mathcal{M})$, with the quantum discord being defined as the difference between the uncertainty about the system for an observer possessing a quantum $\mathcal{M}_{Q}$ and a classical $\mathcal{M}_{C}$ memory, i.e., $\mathcal{D}(\mathcal{S}:\mathcal{M})=S(\mathcal{S}:\mathcal{M}_{Q})-S(\mathcal{S}:\mathcal{M}_{C})$ \cite{Vlatko1} .

The role of information in thermodynamics has been largely discussed in recent literature, with the main focus on the information formulation of thermodynamics (see \cite{VlatkoR} and, for a quantum formulation of thermodynamics, see \cite{QT}). In this direction, a single-mode photo-Carnot engine was considered \cite{Lutz}. The efficiency of the engine was related to the quantum discord and it was demonstrated that it can be larger than that of a purely classical engine, showing that quantum correlations could be an important ingredient in thermodynamics \cite{Lutz}. This result can be viewed as another thermodynamical interpretation of QD.

\subsection{Thermodynamics and the quantum information deficit}

A review on the paradigm concerning the connections between quantum information theory and thermodynamics was presented in \cite{HorodeckiPRA}. Information, in the form of pure quantum states, can be used to extract work from a heat reservoir. In general, by using a system in the state $\rho$ and in contact with a heat bath at temperature $T$, one can draw an amount 
\begin{equation}
W=kT\mathscr{I}(\rho)
\end{equation}
of work, where
\begin{equation}
\mathscr{I}(\rho)=\log d-S(\rho)
\end{equation}
is a measure of information in $\rho$ and $d$ is the dimension of the system's state space ($d=\dim\mathcal{H}$). The quantity $\mathscr{I}(\rho)$ determines the optimal rate of transitions between states under noise operations (NO: unitary transformations, partial trace, and addition of ancillary systems in maximal mixed states). That is, given $n$ copies of $\rho$ one can distil $n\mathscr{I}(\rho)$ pure qubits. Conversely one can take $n\mathscr{I}(\rho)$ pure states and produced $n$ copies of $\rho$, using NO (see \cite{HorodeckiPRA} and references therein for more details about the properties of $\mathscr{I}$).

Local information is a resource and the sound measure of information $\mathscr{I}$ can be used, with actions restricted to closed LOCC operations (CLOCC: unitary transformations and sending systems down a complete dephasing channel), to study how many product pure states (e.g., $|0^{\otimes n}\rangle\otimes|0^{\otimes m}\rangle$) Alice and Bob can distil from a joint state $\rho_{AB}$, in the distant laboratory paradigm. The maximal amount of local information that they can extract in this way is called localizable information and is defined as
\begin{eqnarray}
\mathscr{I}_{l}(\rho_{AB}) &\equiv &\max_{\Lambda\in CLOCC}[\mathscr{I}(\tilde{\rho}_{A})+\mathscr{I}(\tilde{\rho}_{B})]\nonumber\\
&=&\log d_{AB}-\min_{\Lambda\in CLOCC}[S(\tilde{\rho}_{A})+S(\tilde{\rho}_{B})],
\end{eqnarray}
with $d_{AB}=\dim\mathcal{H}_{AB}$ and the extremization is taken over the CLOCC operations. Besides $\tilde{\rho}_{A}$ and $\tilde{\rho}_{B}$ are the reduced states of $\tilde{\rho}_{AB}=\Lambda(\rho_{AB})$. The total information in $\rho_{AB}$, 
\begin{equation}
\mathscr{I}(\rho_{AB})=\log d_{AB}-S(\rho_{AB}),
\end{equation}
can be used to obtain the information that cannot be localized via CLOCC:
\begin{eqnarray}
\Delta(\rho_{AB}) &\equiv &\mathscr{I}(\rho_{AB})-\mathscr{I}_{l}(\rho_{AB}) \nonumber\\
&=&\min_{\Lambda\in CLOCC}[S(\tilde{\rho}_{A})+S(\tilde{\rho}_{B})]-S(\rho_{AB}).
\end{eqnarray}
This quantity was named quantum information deficit (or quantum deficit or work deficit) and can be regarded as a measure of the quantumness of correlations \cite{HorodeckiPRA}.

The information that Alice and Bob can extract acting locally, without using a classical channel to communicate, reads
\begin{eqnarray}
\mathscr{I}_{lo}(\rho_{AB})&=&\mathscr{I}(\rho_{A})+\mathscr{I}(\rho_{B})\nonumber\\
&=&\log d_{AB}-S(\rho_{A})-S(\rho_{B}).
\end{eqnarray}
Thus the classical information deficit:
\begin{eqnarray}
\Delta_{c}(\rho_{AB}) &\equiv &\mathscr{I}_{l}(\rho_{AB})-\mathscr{I}_{lo}(\rho_{AB}) \nonumber\\
&=&S(\rho_{A})+S(\rho_{B}) \nonumber\\
&-& \min_{\Lambda\in CLOCC}[S(\tilde{\rho}_{A})+S(\tilde{\rho}_{B})],
\end{eqnarray}
can be seen as a measure of classical correlation, since it tells us how much more information can be obtained by exploiting additional correlations in $\rho_{AB}$ through a classical channel. For more details of this and related quantities we refer the reader to Ref. \cite{HorodeckiPRA}.

\subsection{Information theoretic interpretations of QD}

Let us now turn our attention to interpretations of QD from the point of view of asymptotic information processing tasks, the so-called information theoretic interpretations. This kind of physical ground is relevant for the consideration of QD as a measure of quantum correlations or as a presumed resource for quantum information tasks. In two articles recently published \cite{Cavalcanti,Datta}, it was given, independently, such an interpretation, by considering the state merging protocol \cite{SMP}. Given an unknown quantum state distributed over two systems, the quantum communication needed to transfer the full state to one of the systems is called partial quantum information. The quantum state merging protocol was introduced to optimally transfer quantum partial information \cite{SMP}. The protocol is based on prior information, shared between both partners, as measured by the conditional entropy. It is important to note that, in this section, when we say conditional entropy we are referring to a direct extension of the classical conditional entropy, which means that we are considering $H\left(A|B\right)\rightarrow S\left(A|B\right)=S\left(A,B\right)-S\left(B\right)$. Note that if there are no correlations between both partners, the conditional entropy is just equal to the entropy of one of the partners. Therefore, the state merging protocol takes advantage of the previous correlations between both partners to reduce the communication cost. For the classical case, the conditional entropy is always positive, implying a positive communication cost. However, in the quantum domain, this quantity may become negative, and in that fact lies the core of the protocol as well as the information-theoretic interpretation of QD. It is important to note here that the protocol as well as the interpretations of QD based on it must be taken in the asymptotic limit of an infinity number of copies. In this case, we can define the regularized discord as 
\begin{equation}
\bar{\mathcal{D}}(\rho_{AB})=\lim_{n\rightarrow\infty}\frac{\mathcal{D}(\rho_{AB}^{\otimes n})}{n}.
\end{equation}
In the remaining of this section we will be referring, unless stated otherwise, to this generalized definition of QD.

Cavalcanti \textit{et al.} \cite{Cavalcanti} considered the state merging protocol as follows: A pure state $\Psi_{ABC}$ is shared between the partners $A$, $B$, and $C$. The aim of the protocol is that $A$ transfers her part to $B$ by using classical communication and shared entanglement between them while maintaining the coherence with the reference state $C$. If the partial information is positive (as it always is in the classical case) $A$ needs to send this number of qubits to $B$. On the other hand, if the partial information is negative (which is permitted in the quantum domain), $A$ and $B$ instead acquire the corresponding potential for future quantum communication \cite{SMP}. In Ref. \cite{Cavalcanti} it was shown that the entanglement consumption in the protocol is equal to the QD between $A$ and $C$, with measurements on $C$. An immediate consequence of these results is an interpretation of the asymmetric character of QD. It is a measure of the difference in the resources needed for $A$ and $C$ to send their parts of the initial state to $B$, through the state merging protocol. As a by-product, they have also noted that the quantum discord may be regarded as an indicator of the direction in which more classical information can be sent through dense coding.

In a related article Madhok and Datta \cite{Datta}, using the strong sub-additivity property of quantum entropy, $S\left(A\vert B,C\right)\leq S\left(A\vert B\right)$, provided another interpretation of QD. This property tells us that having more prior information makes the state merging protocol cheaper. On the other hand, the loss of information will increase the communication cost of the protocol. What the authors in Ref. \cite{Datta} have shown is that the minimum increase of communication cost due to all possible measurements that can be performed on $B$ is given by the QD between $A$ and $B$.

It is worthwhile to note that the state merging protocol is inherently asymmetric (due to the definition of the conditional entropy), so the asymmetry of QD has a direct interpretation as well.

The fact that QD can have both thermodynamic and information-theoretic interpretations, as discussed above, touches in the important relationship between these two (assumed) independent areas, a relation noted years before by Brillouin \cite{Brillouin}. This interesting research field has attracted much attention in the recent literature and certainly deserves further investigations.

\section{Measures of correlations}
\label{MC}

As stated in the Introduction, QD does not seems to be the definitive word for the quantification of non-classical correlations in a general way. This section explores some of the proposed measures of such correlations, others than entanglement, as well as for their classical counterparts. As we shall see, we do not have a closed theory for the characterization and quantification of such correlations. Let us start with what we think that broadly defines quantum correlations, i.e., the impossibility of locally broadcasting it.

\emph{No-Local-Broadcasting Theorem} --- M Piani, P. Horodecki and R. Horodecki have introduced a theorem providing an operational classification of multipartite classical correlations and, hence, for its quantum counterpart \cite{Piani}. Their theorem is based on the impossibility of locally broadcasting, i.e., on the impossibility of locally distributing correlations in order to have many copies of the original state. Purely classical correlations are those permitting such distribution.

For the bipartite case they proved that the correlations (measured by the mutual information) that can be totally transferred from the quantum to the classical world are those present in a state of the form 
\begin{equation}
\rho_{AB}=\sum_{i,j}p_{ij}|i\rangle\langle i|\otimes|j\rangle\langle j|.
\label{CC}
\end{equation}
These are the so-called classical-classical states (CC). In this equation, $\{|i\rangle\}$ and $\{|j\rangle\}$ are orthonormal basis for the subsystems $A$ and $B$, respectively, and $\left\{ p_{ij}\right\} $ is a joint probability distribution. As a consequence, if a given state cannot be cast in the form (\ref{CC}) it must contain non-classical correlations. They then extended this definition to the multipartite case. We note here a close resemblance between this definition and the one regarding the separability problem. Although this result does not really quantify the amount of quantum correlations in a given state, it clearly indicates its presence or not. The authors then go further by proposing a measure for such correlations, in a clear analogy with QD. The proposed quantifier for non-classicality, for the bipartite case, is given by 
\begin{equation}
Q\left(\rho_{AB}\right)\equiv I\left(\rho_{AB}\right)-I_{CC}\left(\rho_{AB}\right)
\label{QDH}
\end{equation}
with $I\left(\rho\right)$ being the mutual information and 
\begin{equation}
I_{CC}\left(\rho\right)=\max_{\{\Pi_{i}^{A},\Pi_{j}^{B}\}}I\left(\sum_{i,j}\left(\Pi_{i}^{A}\otimes\Pi_{j}^{B}\right)\rho_{AB}\left(\Pi_{i}^{A}\otimes\Pi_{j}^{B}\right)^{\dagger}\right)
\label{ClassicalCorr}
\end{equation}
the CC mutual information of state $\rho$. $\{\Pi_{i}^{X}\}$ is a POVM set for partition $X$. The extension of this quantity for the multipartite case depends on the definition of mutual information, which can be obtained by means of a distance in the probability space \cite{Modi}.

Due to the symmetry of Eq. (\ref{CC}) (it is invariant under permutation of $A$ and $B$) we cannot directly translate the interpretations of quantum discord of the last section, which are based on the asymmetric state merging protocol, to the quantifier proposed in Eq. (\ref{QDH}). However, this quantity already has an information theoretic interpretation in terms of the impossibility of locally distributing correlations. This observation leaves open the question about the existence of a thermodynamic interpretation for this quantity.

Equation (\ref{ClassicalCorr}) can be regarded as a symmetric version of QD. The symmetric aspects of correlations quantifiers was discussed in Refs. \cite{LuoOC,Maziero2}. Employing orthonormal projective measures instead of the more general POVMs, an analytical expression for this quantity was also provided in Refs. \cite{LuoOC,Maziero2}, in the case of Bell-diagonal states.

\emph{Thermodynamic Approach} --- Based on the fact that information can be used to extract work from a heat bath, the authors in Ref. \cite{Oppenheim} (see also Sec. \ref{TID} and Ref. \cite{HorodeckiPRA}) have proposed a thermodynamic quantifier for quantum correlations. Considering a scenario in which Alice and Bob share a correlated state they defined the \emph{work deficit} $\Delta$ as the difference in the amount of work that can be extracted by Alice and Bob through CLOCC and by one of the partners holding the whole state, i.e., the work that cannot be locally extracted. In other words, $\Delta$ is a measure of the difference between work that can be extracted in the case where the information is distributed and in the case where it is localized. For the case of pure states, $\Delta$ is exactly equal to the amount of distillable entanglement, while for the general case, it is equal to the amount of quantum correlations (including entanglement) shared by the two partners \cite{Oppenheim}. $\Delta$ is zero only for CC states. This beautiful result reveals, once more, the connection between the thermodynamic work and information, linking them in a measure of correlations.

\emph{Geometric Measure of QD} --- In an interesting article Modi and co-authors \cite{Modi} (see also \cite{Modi2}) provided a geometrical view for measuring quantum correlations. The idea, based on a distance in a probability space, is as follows: A quantifier for a property is given by the distance, as measured by the relative entropy, from a given state (with the property) to the closest state without the desired property \cite{Modi}. The quantum relative entropy between the states $\rho$ and $\sigma$ is given by 
\begin{equation}
S\left(\rho\Vert\sigma\right)\equiv\mbox{Tr}\left[\rho\left(\mbox{log}\rho-\log\sigma\right)\right],
\end{equation}
and it is a measure of the distance, in state space, between these two states. It is important to observe that, although the relative entropy is taken as a distance, it does not satisfy all the properties of a real distance measure; for instance, it is not symmetric.

Given a certain state $\rho$ and the set $\Sigma$ of all separable states, the entanglement is quantified by 
\begin{equation}
E_{R}\left(\rho\right)=\min_{\sigma\in\Sigma}S\left(\rho\Vert\sigma\right),
\end{equation}
while the quantum discord (also named relative entropy of discord) assumes the form 
\begin{equation}
Q_{R}\left(\rho\right)=\min_{\chi\in\Xi}S\left(\rho\Vert\chi\right),
\end{equation}
with $\Xi$ being the set of CC states {[}defined by Eq. (\ref{CC}){]}.

Another quantifier introduced in Ref. \cite{Modi} was the so-called \emph{quantum dissonance}, which, for the state $\rho$, is measured by the minimal relative entropy between the closest separable (denoted as $\sigma$) and the classical states 
\begin{equation}
D_{R}\left(\rho\right)=\min_{\kappa\in\Xi}S\left(\sigma\Vert\kappa\right).
\end{equation}
This is a measure of quantum correlations in separable states, which is analogous to quantum discord, but excludes entanglement. Note that all these measures are valid for whatever the dimension of the system, as well as for the multipartite case.

Another interesting approach was given in Ref. \cite{Dakic}. Having provided a condition for a state to have null QD, the authors proposed the following expression to quantify it 
\begin{equation}
Q_{d}\left(\rho\right)=\min_{\zeta\in\Omega}\Vert\rho-\zeta\Vert^{2}=\mbox{Tr}\left(\rho-\zeta\right)^{2},\label{GDQ}
\end{equation}
with the minimum taken over the set $\Omega$ of zero discord states. And $\Vert\cdot\Vert^{2}$ is the square Hilbert-Schmidt norm. A related dual quantity, named measurement-induced nonlocality, was proposed and studied in Ref. \cite{LuoMIN}.

\emph{Measurement-Induced Disturbance} --- All the above mentioned quantifiers are based on a very difficult extremization procedure involving the set of POVMs. Walking through a different path, Luo has proposed another correlation quantifier based on the disturbance that the measurement processes causes in a system \cite{Lou2}. This quantity, called measurement-induced disturbance (MID), is defined as the difference between the quantum mutual information of the state, $\rho_{AB}$, and that one of the completely dephased state, $\chi_{AB}$
\begin{equation}
MID\left(\rho_{AB}\right)=I\left(\rho_{AB}\right)-I\left(\chi_{AB}\right).
\label{MID}
\end{equation}
As dephasing takes place in the marginal basis, the marginal states are left unchanged. The suitability of MID and its relation to QD are discussed in Ref. \cite{AdessoMID}. It is worthwhile to mention that all quantum correlation quantifiers presented above vanish for the CC states in Eq. (\ref{CC}).

Now let us consider a multipartite system in the state $\rho$. We can quantify its (total) correlations using the distance between $\rho$ and its marginals states in the product form:
\begin{equation}
I(\rho)\equiv S(\rho\Vert\rho_{1}\otimes\rho_{2}\otimes\cdots\otimes\rho_{n})=\sum_{s=1}^{n}S(\rho_{s})-S(\rho),
\end{equation}
where $\rho_{s}$ is the reduced state of subsystem $s$, obtained via the partial trace operation. The state obtained from $\rho$ through local von Neumann measurements reads
\begin{eqnarray}
\tilde{\rho} &=&\sum_{i_{1}=1}^{\dim\mathcal{H}_{1}}\sum_{i_{2}=1}^{\dim\mathcal{H}_{2}}\cdots\sum_{i_{n}=1}^{\dim\mathcal{H}_{n}}\left(\Pi_{i_{1}}\otimes\Pi_{i_{2}}\otimes\cdots\otimes\Pi_{i_{n}}\right)\nonumber\\
&\times &\rho\left(\Pi_{i_{1}}\otimes\Pi_{i_{2}}\otimes\cdots\otimes\Pi_{i_{n}}\right).
\end{eqnarray}
Analogously, the correlation in $\tilde{\rho}$ (which can be interpreted as being the classical correlation in $\rho$ because $\tilde{\rho}$ can be locally broadcast) is given by
\begin{eqnarray}
I(\tilde{\rho}) &\equiv & \max_{\{\Pi_{i_{s}}\}_{i_{s}=1}^{\dim\mathcal{H}_{s}}}S(\tilde{\rho}\Vert\tilde{\rho}_{1}\otimes\tilde{\rho}_{2}\otimes\cdots\otimes\tilde{\rho}_{n}) \nonumber\\
&=& \max_{\{\Pi_{i_{s}}\}_{i_{s}=1}^{\dim\mathcal{H}_{s}}}\sum_{s=1}^{n}S(\tilde{\rho}_{s})-S(\tilde{\rho}),
\label{MCC}
\end{eqnarray}
It is natural to consider the difference \cite{RM}
\begin{eqnarray}
Q(\rho) & \equiv & I(\rho)-I(\tilde{\rho})\nonumber\\
&=&\left[S(\tilde{\rho})-S(\rho)\right]-\sum_{i=1}^{n}\left[S(\tilde{\rho}_{s})-S(\rho_{s})\right]
\label{MQC}
\end{eqnarray}
as a measure of the quantum correlation present in $\rho$, since it quantifies the disturbance in the system due to local measurements. As the von Neumann entropy is non-decreasing under non-selective projective measurements, we identify the two non-negative terms 
\begin{equation}
Q_{nl}(\rho)\equiv S(\tilde{\rho})-S(\rho)
\end{equation}
 and 
\begin{equation}
Q_{l}(\rho)\equiv\sum_{s=1}^{n}\left[S(\tilde{\rho}_{s})-S(\rho_{s})\right]
\end{equation}
as the non-local and local disturbances, respectively. Since our interest here is to measure quantum correlations, we propose that the local measurements $\{\Pi_{i_{s}}\}_{i_{s}=1}^{\dim\mathcal{H}_{s}}$ should be those that maximize (\ref{MCC}).

\emph{Quantum Discord for Continuous Variables} --- All the quantifiers discussed so far were studied considering only the case of discrete variables (with finite dimension), mainly due to the fact that all of them are based on an extremization process that becomes intractable in the case of continuous variables. However, two independent works have extended the QD concept to the continuous variable scenario \cite{Giorda,Adesso2} (for a review about quantum information in continuous variables see \cite{Braunstein}).

In Ref. \cite{Giorda}, focusing in the bipartite case described by two-mode Gaussian states, and within the domain of generalized Gaussian measurement \cite{Cirac}, the authors defined what they called Gaussian discord. They also proved that this quantity is invariant under local unitary operations and is zero only for product states. In Ref. \cite{Adesso2} the authors have computed the QD, as well as its classical counterpart, for all two-mode Gaussian states.

Another interesting article discussing the continuous variables problem studied, instead of QD, the measurement-induced disturbance, shown in Eq. (\ref{MID}), where the two-mode Gaussian state was analysed in some details\cite{Adesso3}.

The relation among the quantifiers for entanglement, total correlation, classical correlation, and quantum discord will not be treated here. The reader can find more details about this issues in Refs. \cite{BrussDE,PianiDE,GharibianED,FanchiniDEoF,CornelioDEi,LiCvS,LiTvQ,LuoEnt,HorodeckiPRA,Pankowski,Qasimi}.

As stated in the introduction we do not have yet a closed theory for the description of the so called non-classical correlations. The developments in this direction lead to the appearance of several measures to quantify (and also characterize) these correlations, but none of them seems to be definitive. Beyond the quantities discussed in this section, we can find many other proposals in literature (see, for example, the papers in Ref. \cite{Pei}).

Trying to fill this gap, C. H. Bennett and co-workers stated three postulates, based on reasonable physical assumptions, which each measure or indicator of genuine multipartite correlation quantifiers should satisfy \cite{Bennett}. The postulates, that apply to correlations in general (including multipartite entanglement or classical correlations) are stated as follows:
\begin{itemize}
\item \emph{First Postulate} --- {}``If an $n$-partite state does not have genuine $n$-partite correlations and one adds a party in a product state, then the resulting $\left(n+1\right)$-partite state does not have genuine $n$-partite correlations.\textquotedblright{}
\item \emph{Second Postulate} --- {}``If an $n$-partite state does not have genuine $n$-partite correlations, then local operations and unanimous post-selection (which mathematically correspond to the operation $\Lambda_{1}\otimes\Lambda_{2}\otimes\cdot\cdot\cdot\otimes\Lambda_{n}$, where $n$ is the number of parties and each $\Lambda_{i}$ is a trace non-increasing operation acting on the $i$-th party's subsystem) cannot generate genuine $n$-partite correlations.\textquotedblright{}
\item \emph{Third Postulate} --- {}``If an $n$-partite state does not have genuine $n$-partite correlations, then if one party splits his subsystem into two parts, keeping one part for himself and using the other to create a new $\left(n+1\right)$-st subsystem, then the resulting $\left(n+1\right)$-partite state does not have genuine $\left(n+1\right)$-partite correlations.\textquotedblright{}
\end{itemize}
The authors then proceed by proposing a definition of genuine $n$- and $k$-partite correlations that satisfy all three postulates \cite{Bennett}:
\begin{itemize}
\item \emph{Definition 1} --- A state of $n$ particles has genuine $n$-partite correlations if it is non-product in every bipartite cut.
\item \emph{Definition 2} --- A state of $n$ particles has genuine $k$-partite correlations if there exists a $k$-particle subset whose reduced state has genuine $k$-partite correlations (according to \emph{Definition 1}). 
\end{itemize}
Here we note that one of the consequences of the postulates stated above is that the existence of genuine $n$-partite quantum correlations without genuine $n$-partite classical correlations is not justified \cite{Bennett}. The authors then proposed a measure of genuine multipartite correlations based on work that can be drawn from local environments by means of a multipartite state.

As we can see, there is a long way yet to be traversed in search of a complete theory to describe, both qualitatively and quantitatively, these correlations.

\subsection{Analytical results for quantum correlations}

\subsubsection{Quantum discord}

As mentioned before, most quantum correlation quantifiers are based on extremization procedures that constitute a very difficult problem, even numerically. Here we review some advances concerning analytical expressions for quantum correlations. Let us first consider the case of two-qubit states with maximal mixed marginals (known as Bell-diagonal states):
\begin{equation}
\rho_{bd}=\frac{1}{4}\left(\mathbf{1}_{AB}+\sum_{i=1}^{3}c_{i}\sigma_{i}^{A}\otimes\sigma_{i}^{B}\right),
\label{BDStates}
\end{equation}
where $\mathbf{1}_{AB}$ is the identity matrix, $\{\sigma_{i}\}_{i=1}^{3}$ are the Pauli spin matrices, and $\{c_{i}\}_{i=1}^{3}$ are real parameters constrained such that $\rho_{bd}$ is a valid density matrix. For this class of states Luo solved the optimization problem and obtained the Henderson-Vedral classical correlation as \cite{Lou}:
\begin{equation}
\mathcal{C}(\rho_{bd})=\frac{1}{2}\sum_{i=0}^{1}\left[1+(-1)^{i}c\right]\log\left[1+(-1)^{i}c\right],
\label{HVCC.BDS}
\end{equation}
with $c=\max(|c_{1}|,|c_{2}|,|c_{3}|)$. The quantum discord is thus given by
\begin{equation}
\mathcal{D}(\rho_{bd})=\sum_{i,j=0}^{1}\lambda_{ij}\log4\lambda_{ij}-\mathcal{C}(\rho_{bd}),
\label{QD.BDS}
\end{equation}
where 
\begin{equation}
\lambda_{ij}=\left[1+(-1)^{i}c_{1}-(-1)^{i+j}c_{2}+(-1)^{j}c_{3}\right]/4
\label{EigBDS}
\end{equation}
are the eigenvalues of $\rho_{bd}$. Also for Bell-diagonal states, it was shown that the formula for the symmetric quantum discord {[}Eq. (\ref{QDH}) with local von Neumann measurements instead of POVM{]} coincides with that in Eq. (\ref{QD.BDS}) for quantum discord \cite{LuoOC,Maziero2}.

A general two-qubit state can be parametrized as (see e.g. \cite{Lou} and references therein) 
\begin{eqnarray}
\rho_{AB} &=& \frac{\mathbf{1}_{AB}}{4} + \vec{x}\cdot\vec{\sigma}^{A}\otimes\frac{\mathbf{1}_{B}}{4} +\frac{\mathbf{1}_{A}}{4}\otimes\vec{y}\cdot\vec{\sigma}^{B} \nonumber\\
&+&\frac{1}{4}\sum_{i,j=1}^{3}t_{ij}\sigma_{i}^{A}\otimes\sigma_{j}^{B},
\label{2qubit}
\end{eqnarray}
where $\vec{x}$ and $\vec{y}$ are the Block vectors for the subsystem $A$ and $B$, respectively, $\vec{\sigma}^{A}=(\sigma_{1}^{A},\sigma_{2}^{A},\sigma_{3}^{A})$, and $\{t_{ij}\}_{i,j=1}^{3}$ are the real elements of the correlation matrix $T$. By definition classical and quantum correlations are invariant under local unitary operations. Thus it is useful to put state (\ref{2qubit}) into a simpler for:
\begin{eqnarray}
\tilde{\rho}_{AB} &=& \frac{\mathbf{1}_{AB}}{4} + \vec{a}\cdot\vec{\sigma}^{A}\otimes\frac{\mathbf{1}_{B}}{4} + \frac{\mathbf{1}_{A}}{4}\otimes\vec{b}\cdot\vec{\sigma}^{B} \nonumber\\
&+& \frac{1}{4}\sum_{i=1}^{3}c_{i}\sigma_{i}^{A}\otimes\sigma_{i}^{B}.
\label{2qubitNF}
\end{eqnarray}
where $\tilde{\rho}_{AB}$ is obtained from $\rho_{AB}$ through local unitary operations and, therefore, it follows that $\mathcal{D}(\tilde{\rho}_{AB})=\mathcal{D}(\rho_{AB})$. In Ref. \cite{Adesso}, Girolami and Adesso regarded the normal form (\ref{2qubitNF}) and obtained a nice algorithm to compute the quantum discord for any state of a two-qubit system. They showed that the measurement direction $(\theta,\varphi)$ attaining the maximum in Eq. (\ref{Dis}) can be obtained by solving the following set of transcendental equations:
\begin{equation}
\lambda_{0}^{-}=\frac{\left(\lambda_{1}^{+}/\lambda_{1}^{-}\right)^{\alpha/\beta}}{1+\left(\lambda_{1}^{+}/\lambda_{1}^{-}\right)^{\alpha/\beta}}\mbox{; }\lambda_{1}^{-}=\lambda_{0}^{-}\left(\frac{\lambda_{0}^{+}}{\lambda_{0}^{-}}\right)^{\frac{\alpha+\beta+\gamma}{2\alpha}},
\end{equation}
where 
\begin{eqnarray}
&\lambda_{0}^{\pm}=\frac{1}{2}\left(1\pm\frac{|\vec{m}_{-}|}{1-\vec{b}\cdot\vec{X}}\right)\nonumber\\
&\lambda_{1}^{\pm}=\frac{1}{2}\left(1\pm\frac{|\vec{m}_{+}|}{1+\vec{b}\cdot\vec{X}}\right)
\end{eqnarray}
are the eigenvalues of $\rho_{A}^{j}$ ($j=0,1$) and
\begin{eqnarray}
&\alpha=\det\begin{bmatrix}\frac{\partial\left(\vec{b}\cdot\vec{X}\right)}{\partial\theta} & \frac{\partial\left(\vec{b}\cdot\vec{X}\right)}{\partial\varphi}\\
\frac{\partial|\vec{m}_{+}|}{\partial\theta} & \frac{\partial|\vec{m}_{+}|}{\partial\varphi}
\end{bmatrix}\nonumber\\
&\beta=\det\begin{bmatrix}\frac{\partial\left(\vec{b}\cdot\vec{X}\right)}{\partial\theta} & \frac{\partial\left(\vec{b}\cdot\vec{X}\right)}{\partial\varphi}\\
\frac{\partial|\vec{m}_{-}|}{\partial\theta} & \frac{\partial|\vec{m}_{-}|}{\partial\varphi}
\end{bmatrix}\nonumber\\
&\gamma=\det\begin{bmatrix}\frac{\partial|\vec{m}_{+}|}{\partial\theta} & \frac{\partial|\vec{m}_{+}|}{\partial\varphi}\\
\frac{\partial|\vec{m}_{-}|}{\partial\theta} & \frac{\partial|\vec{m}_{-}|}{\partial\varphi}
\end{bmatrix},
\end{eqnarray}
with $\vec{m}_{\pm}=(a_{1}\pm c_{1}X_{1},a_{2}\pm c_{2}X_{2},a_{3}\pm c_{3}X_{3})^{t}$ and $\vec{X}=(2\sin\theta\cos\theta\cos\varphi,2\sin\theta\cos\theta\sin\varphi,2\cos^{2}\theta-1)^{t}$, where the superscript $t$ denotes the transpose of vectors or matrices. Further works concerning analytical developments and numerical studies can be found in Refs. \cite{Ali,Fanchini,Lu,ChenX,AdessoMID,Galve,RauGD}.

\subsubsection{Relative entropy of discord and quantum dissonance}

As discussed above, in Ref. \cite{Modi} Modi \emph{et al.} introduced a unified framework to quantify the quantum correlations in any state $\rho$ by means of the relative entropy of $\rho$ and its closest classical state $\chi$ (by classical states we mean those that can be locally broadcast), which is given by 
\begin{equation}
Q_{R}(\rho)=S(\chi)-S(\rho).
\end{equation}
We observe that $Q_{R}$ encompasses all quantum correlations, including those contained in non-separable states. By its turn, the quantum dissonance ($D_{R}$) measures the quantum correlations of separable states. If $\sigma$ is the closest separable state to $\rho$, $D_{R}$ is given by the relative entropy between $\sigma$ and its closest classical state $\kappa$: 
\begin{equation}
D_{R}(\rho)=S(\kappa)-S(\sigma).
\end{equation}
The quantifiers $Q_{R}$ and $D_{R}$ were computed for some classes of states as follows \cite{Modi}:

\emph{Bell-diagonal states.} First let us write the Bell-diagonal states as $\rho=\sum_{i=1}^{4}\lambda_{i}|\Psi_{i}\rangle\langle\Psi_{i}|$, where $\lambda_{i}$ {[}see Eq. (\ref{EigBDS}){]} are ordered in nonincreasing size and $|\Psi_{i}\rangle$ are the four Bell's states. The closest separable state to $\rho$ is $\sigma=\sum_{i=1}^{4}p_{i}|\Psi_{i}\rangle\langle\Psi_{i}|$, with $p_{1}=1/2$ and $p_{i}=\lambda_{i}/[2(1-\lambda_{1})]$ for $i=2,3,4$. The closest classical state for both $\rho$ and $\sigma$ has the form \cite{Modi}:
\begin{eqnarray}
\xi &=&\frac{q}{2}\left(|\Psi_{1}\rangle\langle\Psi_{1}|+|\Psi_{2}\rangle\langle\Psi_{2}|\right)\nonumber\\
&+&\frac{1-q}{2}\left(|\Psi_{3}\rangle\langle\Psi_{3}|+|\Psi_{4}\rangle\langle\Psi_{4}|\right),
\end{eqnarray}
with $q=\lambda_{1}+\lambda_{2}$ for $\xi=\chi$ and $q=p_{1}+p_{2}$ for $\xi=\kappa$.

\emph{Three-qubit W state.} This multipartite pure state is an entangled state and has the form: $|W\rangle=\left(|100\rangle+|010\rangle+|001\rangle\right)/\sqrt{3}$. Its closest separable state is 
\begin{eqnarray}
\sigma &=&\frac{1}{\sqrt{27}}\left(8|000\rangle\langle000|+|111\rangle\langle111| \right. \nonumber\\
&+&\left.12|W\rangle\langle W|+6|\bar{W}\rangle\langle\bar{W}|\right),
\end{eqnarray}
where $|\bar{W}\rangle=\left(|011\rangle+|101\rangle+|110\rangle\right)/\sqrt{3}$. The classically correlated states $\chi$ and $\kappa$ are obtained by dephasing $|\Psi\rangle$ in the $z$-basis and $\sigma$ in the $x$-basis, respectively. Quantitatively we have \cite{Modi} $Q_{R}(|W\rangle)\simeq1.58$ and $D_{R}(|W\rangle)\simeq0.94$. This is an interesting result once it shows that, while for bipartite pure states discord and entanglement coincide, this is not necessarily true for multipartite pure states. There are separable multiparticle pure states that present non-classical correlations.

\emph{Cluster state.} For four qubits the cluster state, which is a resource used in measurement based quantum computation, reads $|C_{4}\rangle=(|0+0+\rangle+|1+1+\rangle+|0-1-\rangle+|1-0-\rangle)/\sqrt{4}$, where $|\pm\rangle=(|0\rangle\pm|1\rangle)/\sqrt{2}$. The closest separable state to $|C_{4}\rangle$ is \cite{Modi} 
\begin{eqnarray}
\sigma &=& \left(|0+0+\rangle\langle0+0+|+|1+1+\rangle\langle1+1+| \right.\nonumber\\
&+& \left. |0-1-\rangle\langle0-1-|+|1-0-\rangle\langle1-0-|\right)/4,
\end{eqnarray}
that is a classically correlated state. Therefore, for this class of states, the relative entropy of entanglement and discord are equal: $E_{R}(|C_{4}\rangle)=Q_{R}(|C_{4}\rangle)=2$ and $D_{R}(|C_{4}\rangle)=0.$

Relative entropy-based measures of quantum correlations were computed for superpositions of N-qubit GHZ and W states in \cite{Parashar}.

\subsubsection{Geometric quantum discord}

In Ref. \cite{Dakic} Daki\'{c} \emph{et al.} introduced the geometric quantum discord {[}Eq. (\ref{GDQ}){]} and obtained an analytical formula for it considering general two-qubit states {[}see Eq. (\ref{2qubit}){]}:
\begin{equation}
Q_{d}(\rho_{AB})=\frac{1}{4}\left(\Vert\vec{x}\Vert^{2}+\Vert T\Vert^{2}-\lambda_{m}\right),
\end{equation}
where $\lambda_{m}$ is the largest eigenvalue of the matrix $\vec{x}\vec{x}^{t}+TT^{t}$. A lower bound for the geometric quantum discord of arbitrary bipartite states was provided in Ref. \cite{Lou1} (see also \cite{HassanGD}). A numerical investigation of the relation between the original quantum discord and its geometric version was carried out in \cite{Batle}.

\subsubsection{Gaussian quantum discord}

As discussed before, the Gaussian quantum discord (GQD) was defined in Ref. \cite{Adesso2} (see also \cite{Giorda}), using generalized Gaussian positive operator valued measurements, and was computed for general two-mode Gaussian states $\rho_{AB}$. This class of states is specified by the covariance matrix (CM) $\mathbf{M}_{c}=\{\mathrm{Tr}(\rho_{AB}\{\hat{R}_{i},\hat{R}_{j}\}_{+})\}$, where $\hat{\mathbf{R}}=\left(\hat{x}_{A},\hat{p}_{A},\hat{x}_{B},\hat{p}_{B}\right)$ is the vector of phase-space operators. All two-mode CMs can be transformed in a standard form
\begin{equation}
\mathbf{M}_{c}=\begin{bmatrix}\mathbf{M}_{1} & \mathbf{M}_{3}\\
\mathbf{M}_{3} & \mathbf{M}_{2}
\end{bmatrix},
\end{equation}
through local unitary operations, with $\mathbf{M}_{1}=\mbox{diag}(a,a)$, $\mathbf{M}_{2}=\mbox{diag}(b,b)$ and $\mathbf{M}_{3}=\mbox{diag}(c,d)$. The CM $\mbox{M}_{c}$ corresponds to a physical state if and only if $A,B\ge1$ and $\nu_{\pm}\ge1$, where the symplectic invariants are defined as $A=\det\mathbf{M}_{1}$, $B=\det\mathbf{M}_{2}$, $C=\det\mathbf{M}_{3}$, $D=\det\mathbf{M}_{c}$ and $\nu_{\pm}^{2}=\left(\delta\pm\sqrt{\delta^{2}-4D}\right)/2$ with $\delta=A+B+2C$. Finally, the Gaussian quantum discord for two-mode Gaussian states is given by \cite{Adesso2}
\begin{equation}
\mathcal{D}(\rho_{AB})=f(\sqrt{B})-f(\nu_{+})-f(\nu_{-})+f(\sqrt{E_{\min}}),
\end{equation}
with
\begin{equation}
f(x)=\sum_{i=0}^{1}(-1)^{i}\frac{x+(-1)^{i}}{2}\log\frac{x+(-1)^{i}}{2}
\end{equation}
and
\begin{eqnarray}
E_{\min} &=&\frac{2C^{2}+(B-1)(D-A)}{(B-1)^{2}} \nonumber\\
&+& \frac{2|C|\sqrt{C^{2}+(B-1)(D-A)}}{(B-1)^{2}}
\end{eqnarray}
if $\ensuremath{(D-AB)^{2}\le C^{2}(B+1)(D+A)}$ and
\begin{eqnarray}
E_{\min} &=& \frac{AB-C^{2}+D}{2B} \nonumber\\
&-& \frac{\sqrt{C^{4}+(D-AB)^{2}-2C^{2}(D+AB)}}{2B}
\end{eqnarray}
otherwise.

\section{Witnessing quantum correlations}

\label{WC}

Despite all the theoretical advances already achieved in this field, its experimental aspects are still in its early stages. As one can note, all the above mentioned measures of quantum correlations require the expensive knowledge of the entire density matrix of the system for their computation, a knowledge usually obtained in the laboratory by means of quantum state tomography. It is also necessary (in general) to perform demanding numerical optimization procedures to compute the amount of non-classical correlations in a given system (as discussed in the last section). A possible way to partially solve this difficulty is by using a correlation witness $W$, a quantity directly accessed in an experiment that can indicate the presence of correlations. Unlike the entanglement-separability paradigm, few witness for detecting such correlations was proposed in the literature \cite{Dakic,Rahimi,Yu,MazieroW,Ma,Laine,BylickaW,ChenW}, and only three experiments were performed \cite{Exp4,Laflamme,Exp6} until now. In this section we discuss the theoretical aspects of correlation witnesses and leave the experimental tests for Sec. \ref{EC}.

In Ref. \cite{Rahimi} it was shown that it is possible to detect non-classical correlations in a single run of a bulk-ensemble experiment. The authors defined a witness map ($W:S\rightarrow\Re$, with $S$ being the state space of the system) possessing the following properties: $(i)$ $W_{\rho}\geq0$ for every state of the form given in Eq. (\ref{CC}) and $(ii)$ there exists non-classically correlated states such that $W_{\rho}<0$. The form of the proposed witness map is given by 
\begin{equation}
W_{\rho}=c-\mbox{Tr}\left(\rho A_{1}\right)\mbox{Tr}\left(\rho A_{2}\right)\mbox{Tr}\left(\rho A_{3}\right)\cdots,
\end{equation}
where the $A_{i}$ are Hermitian operators and $c$ is a constant that must be properly determined. The important difference between this equation and the ones in the entanglement theory is that the present witness (as well as all possible ones) is a \emph{nonlinear} function of the measured operators. This is due to the fact that the non-classical correlated states do not form a convex set \cite{Ferraro,Rahimi}.

Another proposal was reported in Ref. \cite{Yu}, where the authors have provided a single observable witness of an unknown quantum state. Such a protocol demands four copies of the unknown state that may introduce additional experimental tools. To define this witness, let us first define the so-called classical-quantum states. In analogy with Eq. (\ref{CC}) that defines a classical-classical state, a classical-quantum (CQ) one is defined through the relation 
\begin{equation}
\rho_{AB}=\sum_{i}p_{i}|i\rangle\langle i|\otimes\rho_{i}^{B},
\label{CQ}
\end{equation}
where $\{|i\rangle\}$ is an orthonormal basis for the subsystem $A$ and $\rho_{i}^{B}$ are density matrices for partition $B$. This definition is motivated by the fact that, when we compute the QD of this state performing measurements on partition $B$ {[}$\mathcal{D}(A:B)${]} a non-zero value is obtained while, for measurements performed on partition $A$, we always obtain $\mathcal{D}(B:A)=0$. An analogous definition follows for the quantum-classical states, i.e., states such that $\mathcal{D}(A:B)=0$. The no-unilocal-broadcasting theorem for quantum correlations proved in Ref. \cite{LuoSun} characterizes the classical-quantum (quantum-classical) class of states as the only one whose correlations can be locally broadcasted by part $A$ ($B$).

The witness proposed in Ref. \cite{Yu} constitutes a necessary and sufficient condition for the existence of the non-classical correlations in the bipartite, but arbitrarily (finite) dimensional, case (with the QD being computed by measurements on partition $A$). In other words, the witness is zero for states of the form given in Eq. (\ref{CQ}). As a by-product, from the expectation value of the observer that defines the witness one can provide a lower bound for the QD. The witness is defined as 
\begin{equation}
W=\frac{1}{2}\left(X_{A}+X_{A}^{\dagger}\right)\left(V_{13}^{B}V_{24}^{B}-V_{12}^{B}V_{34}^{B}\right),
\end{equation}
where $V_{i,j}^{B}=\sum_{kl}|kl\rangle\langle lk|_{ij}$ acts on the $i$-th and $j$-th copies of partition $B$ of unknown state and $X_{A}=\sum_{klmn}\left\vert klmn\right\rangle \left\langle lmnk\right\vert $ is the cyclic permutation operator acting on partition $A$. The result presented in Ref. \cite{Yu} is that $\mathcal{D}(B:A)=0$ if and only if $\mbox{Tr}\left(W\rho^{\otimes4}\right)=0$.

Considering the two-qubit case, it was reported in Ref. \cite{MazieroW} an experimentally friendly classicality witness for a broad class of such states. The witness is defined by means of the following operators
\begin{equation}
O_{i}=\sigma_{i}^{A}\otimes\sigma_{i}^{B},
\end{equation}
\begin{equation}
O_{4}=\vec{z}\cdot\vec{\sigma}^{A}\otimes\mathbf{1}_{B}+\mathbf{1}_{A}\otimes\vec{w}\cdot\vec{\sigma}^{B},
\end{equation}
with $\sigma_{i}$ being the Pauli matrix in direction $i$ and $\vec{z}$ and $\vec{w}$ are arbitrary real vectors such that $|\vec{z}|=|\vec{w}|=1$. The classicality witness is proposed as 
\begin{equation}
W=\sum_{i=1}^{3}\sum_{j=1+i}^{4}|\langle O_{i}\rangle\langle O_{j}\rangle|.\label{MW}
\end{equation}
The only way to obtain $W=0$ is for a state of the form given in Eq. (\ref{CC}), i.e., a classically correlated state. This condition is only a sufficient one in general, being necessary and sufficient in the case of the Bell-diagonal states. This result was generalized for arbitrary (finite) dimensions in Ref. \cite{Ma}.

Studying the dynamics of the system-environment interaction, in Ref. \cite{Laine} the authors have proposed a witness for the correlations initially shared by system and environment, that can be determined through measurements only on the system. It is interesting to note that, for the determination of the initial correlations, the knowledge of the structure of the environment or of the system-environment interaction is not required.

Using the basis sets $\{A_{i}\}_{i=1}^{d_{A}^{2}}$ and $\{B_{j}\}_{j=1}^{d_{B}^{2}}$ (with $d_{A}=\mbox{dim}\mathcal{H}_{A}$ and $d_{B}=\mbox{dim}\mathcal{H}_{B}$) in the Hilbert-Schmidt spaces of Hermitian operators, one can write any bipartite state as 
\begin{equation}
\rho=\sum_{i,j}c_{ij}A_{i}\otimes B_{j}.
\end{equation}
In Ref. \cite{Dakic}, in addition to the proposal of a geometric measure for QD, the authors introduced a condition for the existence of non-zero quantum discord for any dimensional bipartite states. They showed that if the rank of the correlation matrix $\{c_{ij}\}$ is greater than $d_{A}$($d_{B}$), then the quantum discord obtained by measuring the subsystem $A$($B$) is nonzero. This condition can be seem as a witness for QD and was experimentally verified in Ref. \cite{Laflamme}. Besides the witnesses discussed above other criteria to infer the vanishing of quantum discord can be found in Refs. \cite{Zhu,DattaW}.

\section{Dynamics of correlations under decoherence}
\label{DC}

Besides the characterization and quantification of classical and quantum correlations, another interesting question is the behaviour of these correlations under the action of decoherence. This phenomenon, mainly caused by the injection of noise into the system, arising from its inevitable interaction with the surrounding environment, is responsible for the loss of quantum coherence initially present in the system (see \cite{Decoherence} for a modern treatment on this subject). This is a very important topic not only from a fundamental point of view, but also for practical purposes envisaging the development of quantum information protocols that make use of this kind of correlations. In this section we will discuss the main theoretical developments on this subject, letting the experimental investigation for Sec. \ref{EC}.

\subsection{Markovian dynamics of correlations}

Considering the Markovian dynamics of two-qubit systems, in Ref. \cite{Werlang} the authors showed that, in some cases where the entanglement sudden death \cite{ESD} occurs, the quantum discord only vanishes in the asymptotic limit. It was analysed the case of two non-interacting qubits under the action of three, independent, decoherence channels, i.e., the dephasing, the depolarizing and the generalized amplitude damping channels. Initial conditions that lead to the entanglement sudden death were chosen and the dynamics of QD was investigated \cite{Werlang}. This result can be directly extended to more general cases, remembering that states of zero discord form a set of null measure and that is nowhere dense in the state space \cite{Ferraro}.

Another interesting result concerning decoherence was obtained by Maziero and co-workers \cite{Maziero}. Studying the effect of the Pauli maps (phase, bit-flip and bit-phase-flip channels) on two non-interacting qubits, the authors have predicted the \emph{sudden change} (SC) phenomenon, as well as the immunity against decoherence of correlations, both experimentally confirmed in Refs. \cite{Exp1,Exp2}. The SC phenomenon consists in an abrupt change in the decay rate of the correlations, highly dependent on the geometry of the initial state. Initially it was thought that such phenomenon will appear only under the dynamics of the Pauli maps, however it was found to occur in the presence of a thermal bath \cite{SCthermal,Exp2} and also in a squeezed common reservoir \cite{Maria}. This peculiar behaviour was found again in the case of non-inertial qubits \cite{AliceRob}, being caused by the Unruh effect \cite{Matsas}. These results indicate that the SC could be a dynamic characteristic of the correlations beyond entanglement, extending the ideas presented in Ref. \cite{Maziero}. However, a physical interpretation of such behaviour is still lacking.

On the other hand, considering the immunity against decoherence, it was found that, under certain conditions, the classical correlations can remain unaffected by decoherence, while its quantum counterpart goes to zero \cite{Maziero}. This fact directly leads to an operational measure (without any extremization procedure) of correlations. The quantum correlations of the initial state can be computed through the difference between the initial mutual information and the mutual information of the completely decohered state \cite{Maziero}. Contrary to what happens with SC, this phenomenon does not take place in a thermal environment, as experimentally demonstrated in Ref. \cite{Exp2}.

In a related work, it was found that, not only the classical correlations, but also the quantum ones, can remain, under certain circumstances and for a fixed period of time, unaffected by decoherence (considering only Pauli maps) \cite{Mazzola}. It was discovered that a class of two-qubit states for which the quantum correlations remain unaffected until the transition time, after which they began to decay while the classical correlations become immune to decoherence \cite{Mazzola}. A geometrical picture for the non-analytic behaviour of quantum correlations under decoherence, discussed above, was provided by Lang and Caves in Ref. \cite{Lang} (see also \cite{Fei}).

In a different context, in Ref. \cite{MazieroSR} the authors have investigated the role of the correlations between \emph{system and environment} in their decoherent dynamics. It was regarded a two-qubit system under the action of two independent reservoirs, considering beyond the Pauli maps, also the amplitude damping channel. The two main results reported in this reference are: (i) The decoherence phenomenon may occur without bipartite entanglement between system and environment and (ii) the initial non-classical correlations (presented in the two-qubit system) completely disappear, in some circumstances, being not transferred to the environments \cite{MazieroSR}.

Concerning the continuous variables scenario, the dynamics of the Gaussian QD between two non-interacting modes under the action of a thermal reservoir was analysed \cite{Isar}. The dynamics of QD and the entanglement was compared and, while entanglement suffers sudden death (in certain cases), the QD decays asymptotically to zero, for the situations analysed in this reference.

Common environments acting on bipartite system was considered in Refs. \cite{Auyuanet,Yuan}. In Ref. \cite{Auyuanet}, the authors considered two initially excited qubits, interacting with the modes of the electromagnetic field in the vacuum state. This constitutes a common reservoir for both qubits. It was demonstrated that the quantum and classical correlations are generated before entanglement. They also studied the QD and the MID dynamics \cite{Auyuanet}. In Ref. \cite{Yuan}, the authors have considered an Ohmic thermal environment acting on two non-interacting qubits. It was demonstrated the existence of a stable amplification of QD for identical qubits, while for distinct qubits (with large detuning), the protection of QD from the environment action is possible in some cases. However, such phenomena can occur only before the sudden change time \cite{Maziero}. Once more, we observe that the SC behaviour seems to be an important issue in the decoherence dynamics of the quantum and classical correlations.

All the works discussed so far are dealing with Markovian environments, i.e., environments that do not present memory effects. Next we review some of the developments that treat the dynamics of correlations under the action of non-Markovian environments.

\subsection{Non-Markovian dynamics of correlations}

In Ref. \cite{Wang} the authors have considered the case of two non-interacting qubits under the action of two independent zero temperature non-Markovian thermal environments. Comparing the dynamics of entanglement with that one for QD, it was discovered that while the first presents sudden death, the second only disappears at some times \cite{Wang}. The case of two qubits interacting with a common reservoir was considered in Ref. \cite{Fanchini}, where SC was again observed. The results presented in this last work seems to indicate that, for a general initial conditions, the SC will occur. Two non-interacting harmonic oscillators under the action of independent and common non-Markovian bosonic reservoirs were considered in the context of continuous variables in Ref. \cite{Vasile}. An interesting result reported in this article is that, in the case of a common reservoir, while QD can be created between the oscillators, the entanglement remains absent. In Ref. \cite{Mazzola1} the authors have analysed the case of two qubits interacting with a non-Markovian depolarizing channel. The appearance of multiple sudden change was observed. Similar results were experimentally observed in Ref. \cite{Exp3} (see also Sec. \ref{EC}).

\subsection{Quantum discord and completely positive maps}

Besides the results discussed above, there is an interesting issue linking quantum discord and the evolution of open quantum systems. Rodrguez-Rosario \textit{et al.} proved that if the system-environment initial state ($\rho_{SE}$) is a classical-quantum state, i.e., if the discord obtained by measuring the system is null, then the system's dynamics is described by a completely positive map \cite{Rodriguez}. They also presented examples of separable initial states $\rho_{SE}$, with nonzero discord, for which the dynamical map describing the system's time evolution is not positive \cite{Rodriguez}. This is a striking result concerning quantum discord, because it suggest something fundamental about correlations, i.e., they can influence the dynamics of the system under the action of the environment in a profound way. 

In Ref. \cite{Shabani} two questions were faced: (i) Is the dynamics of the open quantum system equivalent to a map between the initial and the final sates? (ii) Is this map completely positive? The answer to the first question is affirmative, with the map being linear and Hermitian for all initial conditions with the system-environment states belonging to the so called special linear class of states:
\begin{equation}
\rho_{SE}=\sum_{ij}c_{ij}|i\rangle\langle j|\otimes\phi_{ij},
\end{equation}
where $\{\phi_{ij}\}_{i,j=1}^{\dim\mathcal{H}_{S}}:\mathcal{H}_{E}\mapsto\mathcal{H}_{E}$ with $\mathrm{Tr}\phi_{ij}=1$ or $\phi_{ij}=0$ $\forall i,j$, and $\{|i\rangle\}_{i=1}^{\dim\mathcal{H}_{S}}$ is an orthonormal basis for $\mathcal{H}_{S}$. The answer for the second one is negative for initial special linear states presenting non vanishing quantum discord.

Also with concern to fundamental aspects of the system-environment dynamics, quantifiers for the classical and quantum decorrelating capabilities of a quantum operation were introduced and analysed in Ref. \cite{LuoDC}. And it was shown in Ref. \cite{RosarioER} that, under the presence of any interaction with the environment, the quantum entropy rate of a system is zero if and only if the system-environment state has zero quantum discord, with projective measurements of higher rank in the system.

\section{Experimental investigations}
\label{EC}

This section is devoted to the experimental aspects of quantum and classical correlations. Only a few experiments on this issue have been reported up-to-date in literature.

The first experiment in this context was performed to test the conjecture that the quantum correlations, as measured by QD, are the figure of merit for the speed-up in mixed-state quantum computation without or with little entanglement \cite{White}. Using an all-optical architecture, the generation of correlations in the trace estimation DQC1 (Deterministic quantum computation with one qubit) protocol was observed and no entanglement has been found, despite the fact that an exponential speed-up (in comparison with the best known classical protocol) is obtained with this protocol. However, a large amount of QD was generated, indicating that, although fully separable, mixed states can contain non-classical correlations that may be a valuable resource for quantum information technology \cite{White}.

From a more fundamental point of view, in Ref. \cite{Exp1} the authors have confirmed experimentally the sudden change behaviour of correlations, as well as their immunity against some sources of decoherence, both phenomena theoretically predicted in Ref. \cite{Maziero}. Also using an all-optical setup the authors followed the dynamics of correlations under the action of a simulated dephasing environment in a controllable way. The experimental observation showed an excellent agreement with the theory \cite{Exp1}. In this experiment, the environment is a non-dissipative one, i.e., there is no energy exchange between system and environment. Only phase relations are lost. To fit this gap, in Ref. \cite{Exp2} the same phenomena was demonstrated, but in this case, with the presence of a real dephasing channel and also of the amplitude damping channel, responsible for the energy exchange between system and environment. Such phenomena was shown to occur in a liquid state room temperature nuclear magnetic resonance (NMR) experiment \cite{Exp2}. This result indicates that the SC may occur in general scenarios being therefore a general property of quantum correlations.

The decoherence dynamics of correlations under the action of two independent amplitude damping channels (one for each qubit) and a global phase damping one was studied in Ref. \cite{ExpDio}. The experiment was performed in a solid state NMR system at room temperature. A theory to obtain the correlations from the NMR deviation matrix was also developed in this article, and an excellent agreement with the experimental data was obtained \cite{ExpDio}, uncovering thus the quantum nature of correlations in room temperature NMR systems. Another reported experiment concerns the non-Markovian character of the decoherent dynamics of correlations \cite{Exp3}. By means of a controlled dephasing non-Markovian environment, the multiple sudden change behavior was observed.

More recently, it was reported the first experimental observation of a witness for quantum correlations \cite{Exp4}, theoretically proposed in \cite{MazieroW}. Also employing the liquid NMR architecture, the authors of Ref. \cite{Exp4} have measured directly the witness in Eq. (\ref{MW}) with high precision. This result implies that it is possible to determine the quantum nature of correlations by means of only a few local measurements, without the necessity of the expensive tomographic process and also avoiding the extremization procedures involved in the quantification of quantum correlations. A modified version of the witness proposed in Ref. \cite{MazieroW} was presented and also implemented in an optical setup through the measurement of a single observable in \cite{Exp6}.

The condition for non-zero quantum discord introduced in Ref. \cite{Dakic} (based on the rank of the correlation matrix) was employed in the experiment reported in Ref. \cite{Laflamme} in the NMR setup, in order to demonstrate the non-classicality in the DQC1 algorithm using four qubits.

\section{Applications and related effects}
\label{AC}

In this section, we will discuss some applications of these general quantum and classical correlations in different contexts, like quantum computation and biological systems. As we shall see, there is an increasing interest in such studies due to both fundamental and technological reasons.

\subsection{The source of quantum advantage}

It is well accepted today that quantum mechanics can provide us a gain (in the computational time, or in the parameter estimation, for instance) over equivalent classical protocols. However, the source of such a gain, initially thought to be the entanglement, is not so clear \cite{VedralSource,Eastin}. In this subsection we shall present some recent developments in this direction.

In an article published in Physical Review A, Datta and Gharibian \cite{Gharibian} have analysed the role played by quantum correlations in the so-called mixed-state quantum computation, focusing in particular on the DQC1 model, proposed by Knill and Laflamme \cite{KL}. Considering separable states, the authors have found indications that the quantum correlations, as quantified by MID, could be the source of the speed-up in the DQC1 model \cite{Gharibian}. The QD was proposed as being the source of the quantum advantage in the DQC1 model in Ref. \cite{DattaDQC1} (see also \cite{DattaPhD}). Studying how correlations behave in the Deutsch-Jozsa algorithm \cite{DJ}, the authors of Ref. \cite{Chaves} have concluded that entanglement is not the unique signature of efficient quantum computation, and it is not even necessary for protocols which present gain over their classical counterparts \cite{Chaves}. The prisoners dilemma and other quantum games were analysed in Ref. \cite{Nawaz} with similar conclusions. By analysing the mixed-state phase estimation, in Ref. \cite{MSPS} it was found that, although classical correlations do not play an important role in this process, the $\sqrt{N}$ gain is still present when entanglement vanishes, but not the QD. It was shown in \cite{BoixoQL} that quantum discord is responsible for the advantage in the quantum locking protocol, where entanglement is known to play no role. The connection between the mother protocol of quantum information theory and quantum discord was discussed in \cite{MadhokMP}.

Walking in the opposite direction, in Ref. \cite{Terno}, it was shown that the ability to create entanglement is necessary for the execution of bipartite quantum gates. In other words, to execute such a gate, starting with the LOCC should be supplemented by shared entanglement. QD is then related to the failure of the LOCC implementation of the gate \cite{Terno}. A related work, linking QD and entanglement as a resource for quantum computation was presented in Ref. \cite{Marcos} (see also \cite{Dakic,VedralSource}). In 1997, Grover proposed an algorithm using quantum mechanics to obtain a quadratic temporal speed-up in search applications over unsorted data \cite{Grover}. In a recent article the role of classical and quantum correlations was analysed in such scenario \cite{Cui}. It was argued that entanglement works as the indicator of the increasing rate of the success probability of the algorithm \cite{Cui}.

As we can see, despite all the progress, the debate about the source of quantum effectiveness is still an open question.

\subsection{Relativistic effects}

Quantum theory and Relativity are the two main pillars on which rests the modern physics. Therefore, it is natural to ask how the behaviour of quantum correlations is changed when Relativity comes into play. This subsection only deals with relativistic effects on quantum discord. A discussion on the role of Relativity in information theory can be found in Ref. \cite{Peres}.

Contrary to entanglement, there are only a few articles discussing the relativistic aspects of non-classical correlations so far. The first one is due to Datta \cite{DattaR}. The system under consideration is a two, initially entangled, modes of the scalar field. The correlations shared by them were analysed in the case when the observer detecting one of the modes undergoes a constant acceleration, while the other stayed inertial. It was found that while entanglement dies due to the Unruh effect, a considerable amount of quantum discord still survives, even in the limit of infinite acceleration \cite{DattaR}, which is a very interesting result since, in this limit, the accelerated observer should experience an infinite temperature thermal bath \cite{Matsas}.

The dynamics of classical and quantum correlations in a two-qubit system when one of them is uniformly accelerated during a finite amount of proper time was analysed by C\'{e}leri and collaborators in Ref. \cite{AliceRob}. In this case, while the quantum correlations, as measured by symmetric quantum discord \cite{Piani,Maziero2}, is completely destroyed in the limit of infinite acceleration, the classical one remains nonzero \cite{AliceRob}. It is interesting to note here that such correlations exhibit the SC behavior \cite{Maziero} as a function of acceleration (the temperature of the associated Unruh thermal bath experienced by the qubit is directly proportional to its proper acceleration \cite{Matsas}). These results were then extended to the case where the qubits lie in the vicinity of a black hole, with one of them freely falling into the hole (the inertial qubit) while the other staying static (the accelerated one) near the event horizon.

In another interesting article, the case of correlations between modes of Dirac fields in non-inertial frames was considered \cite{Wang}. It was found that the classical correlations decrease as the acceleration increases, which is in sharp contrast with the scalar field case where the classical correlations do not depend on acceleration \cite{AdessoR}. Moreover, the original correlations shared by Alice and Rob (the accelerated Bob) from an inertial perspective are redistributed between all the bipartite modes from a non-inertial perspective \cite{WangR}.

Therefore, the dynamic behaviour of classical and quantum correlations can be very different depending on the system under consideration. This is an issue that certainly deserves more study to permit us to understand its dynamic fundamental aspects. Besides the fundamental questions, relativistic effects on correlations became important for the attempt of constructing global communications systems based on quantum mechanics.

\subsection{Critical systems}

The study of critical systems has attracted the interest of a wide variety of scientists, both from the classical and quantum viewpoints. In general, the main element for such studies is the concept of correlations. Indeed, changes in these correlations strongly affect the observable properties of a many-body system and are responsible, for instance, to quantum phase transitions (QPTs), which is a critical change in the ground state of a quantum system due to level crossings in its energy spectrum (for an introduction on quantum phase transitions see \cite{Sachdev}).

We will discuss here the developments obtained in the context of quantum and classical correlations concepts applied to quantum critical systems. We shall discuss only some of these developments and refer the reader to other references.

Quantum discord was applied to the study of QPTs in the transverse field Ising model 
\begin{equation}
H_{I}=-\sum_{i}\left(\lambda\sigma_{i}^{x}\sigma_{i+1}^{x}+\sigma_{i}^{z}\right)
\end{equation}
 and in the antiferromagnetic \textit{XXZ} spin $1/2$ model 
\begin{equation}
H_{XXZ}=\sum_{\langle ij\rangle}\left(\sigma_{i}^{x}\sigma_{j}^{x}+\sigma_{i}^{y}\sigma_{j}^{y}+\Delta\sigma_{i}^{z}\sigma_{j}^{z}\right),
\end{equation}
with the sum running over the nearest-neighbours $\langle ij\rangle$ \cite{QPT1}. For both models, in the thermodynamic limit, the amount of QD increases close to the critical points, indicating that QD could be a good indicator of the existence of such points in the system, as a qualitative detector. On the other hand, the derivatives of the QD (with respect to the parameter that drives the QPT) provide precise information on the location and on the order of the QPTs \cite{QPT1}. Such analysis was generalized by means of analytical expressions for the correlations in each model and also by discussing their finite-size behaviour in Ref. \cite{Sarandy}. Signatures of the critical behaviour of the system for the cases of first-order, second-order, and infinite order QPTs was found in both classical correlations and QD \cite{Sarandy}.

The pairwise correlation in the two-spin one-dimensional \textit{XYZ} Heisenberg model, 
\begin{eqnarray}
H_{XYZ} &=& B\left(\sigma_{1}^{z}+\sigma_{2}^{z}\right)+J_{x}\sigma_{1}^{x}\sigma_{2}^{x}+J_{y}\sigma_{1}^{y}\sigma_{2}^{y} \nonumber\\
&+&  J_{z}\sigma_{1}^{z}\sigma_{2}^{z},
\end{eqnarray}
at finite temperature was considered in Ref. \cite{Rigolin}. The authors have studied the dependence of QD on the temperature and also on the external magnetic field $B$ acting on both qubits. A very distinct behaviour from entanglement was found and it was conjectured that QD is able to detect a QPT even at finite temperature \cite{Rigolin}.

In Ref. \cite{MazieroQPT} it was reported a study of the QD behaviour in the thermodynamic limit of the anisotropic \textit{XY} spin-$1/2$ chains in a transverse magnetic field 
\begin{eqnarray}
H_{XY} &=& -\sum_{i} \frac{\lambda}{2}\left[\left(1+\gamma\right)\sigma_{i}^{x}\sigma_{i+1}^{x}+\left(1-\gamma\right)\sigma_{i}^{y}\sigma_{i+1}^{y}\right] \nonumber\\
&-& \sum_{i}\sigma_{i}^{z} 
\end{eqnarray}
considering both cases, for zero and finite temperatures. After obtaining analytical expressions for pairwise correlations at any distant sites, it was shown that, at zero temperature, QD can be able to detect a QPT even for situations where entanglement fails in this task. For finite temperatures, the authors showed that QD can increase with temperature in the presence of a magnetic field \cite{MazieroQPT}.

A remarkable property of QD was provided in Ref. \cite{Rigolin2} where, analysing the \textit{XXZ} model, it was shown that, indeed, QD can indicate the critical points associated with QPT. This result is in sharp contrast to entanglement as well as to others thermodynamic quantities.

Another interesting result was recently reported in Ref. \cite{MazieroQPT1}. Pairwise QD was analysed as a function of distance between spins in the transverse field XY chain and in the XXZ chain in the presence of domain walls. It was found a long-range decay of quantum discord, that indicate the critical points in these systems. This behaviour is rather different from entanglement, which is typically short-ranged. Moreover, a clear change in the decay rate of correlations when the system crosses the critical point was predicted \cite{MazieroQPT1}.

Beyond these developments, there are many other articles in recent literature considering the quantum aspects of correlations in critical systems. For instance, the case of the \textit{XY} chain was considered in Ref. \cite{Ciliberti}, the \textit{XX} model in a nonuniform external magnetic field was reported in Ref. \cite{Hassan} and the phenomenon of symmetry breaking in a many body system in Ref. \cite{Tomasello}. A comparison of the QD with the work-deficit was provided in Ref. \cite{Dhar} and the MID in Ref. \cite{ZhangQPT}. The QD and entanglement were computed and compared for some systems whose
ground states can be expressed as matrix product states in \cite{SunMPS}.

All the above results show that QD could play an important role in critical systems from both point of views theoretical and experimental. So, it is a very interesting quantity to be explored in the condensed matter scenario.

\subsection{Biology}

The quantum description of biological molecules and, in particular, the role that quantum mechanics could play in biological processes, have attracted much attention in the recent literature (see, for example, \cite{Bio}). Although very controversial, some developments were obtained, mainly linking entanglement with the efficiency of a given process \cite{NatExp}. The aim of this subsection is to present the study of classical and quantum correlations in such a systems, recently reported in Ref. \cite{CorBio}.

The goal of this program is to identify whether quantum mechanical effects are present in biological systems, and in Ref. \cite{CorBio}, the authors have analysed the role that QD, and its classical counterpart, play in the transfer of an excitation from a chlorosome antenna to a reaction center, by the so-called Fenna-Matthews-Olson protein complex. The numerical simulations of Ref. \cite{CorBio} indicate that a significant fraction of the total correlation in such a process is due to QD in the first picosecond (the relevant time scale for this dynamics). It is important to note that this result was obtained both at cryogenic and physiological temperatures \cite{CorBio}.

Despite very speculative, these works raise many questions about the role played by quantum mechanics in biological systems. Concerning quantum correlations we could ask how the efficiency of a given process scales with quantum discord (or other correlation measure)? Another interesting question is if it is possible to model some molecules, like the DNA, by means of a spin chain and, as a consequence, we could apply the results of the preceding subsection to these systems.

Besides the developments presented here, there are many articles devoted to the study of classical and quantum correlations in several others scenarios. For instance, in Ref. \cite{WangC} the dynamics of QD between two atoms interacting with a cavity field was analysed and the spin-boson model in Ref. \cite{SB}. Applications in the solid state scenario can be found for the case of quantum dots in Ref. \cite{QD}.

\section{Concluding remarks}
\label{CR}

Since the publication of the EPR criticism to quantum mechanics, the concept of classical and non-classical correlations has been greatly modified, as we have seen in the introduction of this article. We have evolved from correlations violating a Bell's inequality to quantum correlations measured, for example, by the quantum discord, passing through entanglement. In this short review, we focused on the non-classical aspects of correlations presented in separable states, as the ones revealed by the quantum discord. We saw that there are several quantifiers for such correlations and none of them seems to be definitive. Despite the development achieved recently in this issue, the question concerning what is quantum in a correlated system, is still a bit open.

Beyond its fundamental aspects, another important and controversial subject was raised together with the birth of the concept of quantum discord-like correlations, the one concerning its application for quantum information processing. The main question in this direction is what is responsible for the advantages (over purely classical systems) provided by quantum theory. Besides some controversial questions there are advances in this direction. For instance, we certainly know that such kind of non-classicality plays an important role in communication, as revealed by the non-local broadcasting \cite{Piani}, and in quantum metrology \cite{MSPS}, among other scenarios. The investigation of the precise role of separable mixed states with non-classical correlations opens a very exiting avenue of research.

The experimental investigation of the behaviour of these correlations has been started only very recently. We have only a few experimental investigations at our disposal, and much more need to be done in order to lead us to a better test and comprehension about the quantum nature of correlations in composed systems.

The application of these correlations to another areas like condensed matter physics and biological systems is in its early stages and a long way is still to be traversed in order to appreciate the true role of these correlations in these different fields of knowledge. We hope that this review will call attention to the importance of discord-like correlations and will stimulate further research on the subject.

\begin{acknowledgments}
The authors acknowledge financial support from UFABC, CAPES, and FAPESP. This work was performed as part of the Brazilian National Institute for Science and Technology of Quantum Information (INCT-IQ). \end{acknowledgments}

\end{document}